\documentclass[final, 5p, times, twocolumn]{elsarticle}

\usepackage{amsmath,amsfonts,amssymb,bm}
\usepackage{color}
\usepackage{dsfont}
\usepackage{graphicx}
\usepackage{multirow}
\usepackage{soul}
\usepackage{CJKutf8}

\usepackage[mathscr,scaled=1.15]{urwchancal}
\DeclareFontFamily{OT1}{pzc}{}
\DeclareFontShape{OT1}{pzc}{m}{it}%
{<-> s * [1.15] pzcmi7t}{}
\DeclareMathAlphabet{\mathpzc}{OT1}{pzc}{m}{it}

\definecolor{purple}{rgb}{0.5,0,0.5}
\definecolor{blue}{rgb}{0.0,0,0.9}
\definecolor{prdblue}{rgb}{0.133,0.118,0.498}
\usepackage[colorlinks=true, pdfstartview=FitV, linkcolor=prdblue, citecolor= prdblue, urlcolor=prdblue]{hyperref}

\biboptions{sort&compress}

\journal{Physics Letters B}

\begin{document}

\begin{CJK}{UTF8}{song}

\begin{frontmatter}



\title{$\,$\\[-7ex]\hspace*{\fill}{\normalsize{\sf\emph{Preprint no}. NJU-INP 052/21}}\\[1ex]
Semileptonic transitions: $B_{(s)} \to \pi(K)$; $D_s \to K$; $D\to \pi, K$; and $K\to \pi$}


\author[NJU,INP]{Zhao-Qian Yao}
\ead{zqyao@smail.nju.edu.cn}

\author[ECT]{Daniele Binosi}
\ead{binosi@ectstar.eu}

\author[NJU,INP]{Zhu-Fang Cui}
\ead{phycui@nju.edu.cn}

\author[NJU,INP]{Craig D. Roberts\corref{cor2}}
\ead{cdroberts@nju.edu.cn}
\cortext[cor2]{Corresponding Author}

\address[NJU]{
School of Physics, Nanjing University, Nanjing, Jiangsu 210093, China}
\address[INP]{
Institute for Nonperturbative Physics, Nanjing University, Nanjing, Jiangsu 210093, China}

\address[ECT]{
European Centre for Theoretical Studies in Nuclear Physics
and Related Areas, Villa Tambosi, Strada delle Tabarelle 286, I-38123 Villazzano (TN), Italy}

\begin{abstract}
Continuum Schwinger function methods for the strong-interaction bound-state problem are used to arrive at a unified set of parameter-free predictions for the semileptonic $K\to \pi$, $D\to \pi, K$ and $D_s \to K$, $B_{(s)} \to \pi(K)$ transition form factors and the associated branching fractions.  The form factors are a leading source of uncertainty in all such calculations: our results agree quantitatively with available data and provide benchmarks for the hitherto unmeasured $D_s\to K^0$, $\bar B_s \to K^+$ form factors.  The analysis delivers a value of $|V_{cs}| = 0.974(10)$ and also predictions for all branching fraction ratios in the pseudoscalar meson sector that can be used to test lepton flavour universality.  Quantitative comparisons are provided between extant theory and the recent measurement of ${\mathpzc B}_{B_s^0\to K^- \mu^+ \nu_\mu}$.  Here, further, refined measurements would be useful in moving toward a more accurate value of $|V_{ub}|$.
\\[1ex]
\leftline{2021 November 10}
\end{abstract}



\begin{keyword}
heavy-quark mesons \sep
semileptonic decays \sep
CKM matrix elements \sep
emergence of hadron mass \sep
Nambu-Goldstone bosons \sep
Schwinger function methods
\end{keyword}

\end{frontmatter}
\end{CJK}



\section{Introduction}\label{SecIntro}
\unskip
In the Standard Model (SM), quark current-masses and weak-interaction mixing between quark flavours have a common origin, \emph{viz}.\ couplings to the Higgs boson.  Through many years, much experimental and theoretical effort has been devoted to the precise determination of all parameters involved.  The status of current-mass estimates is described in Ref.\,\cite[Sec.\,59]{Zyla:2020zbs}.  A second source of mass is, however, overlooked in such considerations, \emph{viz}.\ that which is seemingly connected with strong interactions in the SM.  This is emergent hadron mass (EHM), responsible, \emph{inter alia}, for the 1\,GeV scale that characterises all visible matter \cite{Roberts:2020hiw, Roberts:2021nhw}.  Understanding EHM is crucial to explaining the behaviour of Nature's three lightest quarks and all systems in which they feature as the valence degrees-of-freedom.  EHM also influences heavy-quark systems, \emph{e.g}., being involved in producing roughly 20\% of the $B$-meson mass \cite{Xu:2021iwv}.  These things feed into the uncertainties attending upon determination of the elements of the Cabibbo-Kobayashi-Maskawa (CKM) matrix, which parametrises quark flavour mixing.  They must be properly expressed in any approach that attempts to supply a unified explanation of hadron properties.

A poorly known CKM matrix element is that which characterises $b\to u$ transitions: an average of a selection of disparate results yields \cite[Sec.\,75.3]{Zyla:2020zbs} $|V_{ub}| = 0.00382(24)$.  It is typically judged \cite{Gambino:2020jvv} that the most precise determinations will be achieved using theoretically clean processes that involve ground-state hadrons in the final state, such as $\bar B^0 \to \pi^+ \ell^- \nu_\ell$, $\bar B_s^0 \to K^+ \ell^- \nu_\ell$.  Measurements of the former are available \cite{BaBar:2010efp, Belle:2010hep, BaBar:2012thb, Belle:2013hlo} and $B_s^0 \to K^- \mu^+ \nu_\mu$ has recently been observed \cite{LHCb:2020ist}.
Here, ``clean'' means, for instance, that the processes involve only the vector component of the weak interaction vertex and the initial and final states are systems constituted from just two valence degrees-of-freedom.  There are, nevertheless, obstacles for theory to overcome: (\emph{i}) the huge disparity between the masses of the initial and final states; and (\emph{ii}) the Nambu-Goldstone boson character of pions and kaons.  The latter is an especially striking expression of EHM.  Today, no framework with a traceable connection to quantum chromodynamics (QCD) can simultaneously and directly surmount both these challenges.  Nevertheless, attempts are being made, \emph{e.g}., Refs.\,\cite{Melikhov:2001zv, Faessler:2002ut, Ebert:2003wc, Ball:2004ye, Wu:2006rd, Khodjamirian:2006st, Lu:2007sg, Ivanov:2007cw, Faustov:2013ima, Xiao:2014ana, Wang:2015vgv, Lu:2018cfc, Zhang:2020dla, Gonzalez-Solis:2021awb, Bouchard:2014ypa, Flynn:2015mha, FermilabLattice:2015mwy, FermilabLattice:2019ikx}.
%

We approach the problem using the continuum Schwinger function methods (CSMs) \cite{Eichmann:2016yit, Qin:2020rad} employed successfully to provide a unified explanation for the properties of hadrons with $0-3$ heavy quarks, \emph{viz}.\ from the lightest (almost) Nambu-Goldstone bosons to triply heavy baryons, \emph{e.g}., Refs.\,\cite{Ding:2018xwy, Binosi:2018rht, Qin:2019hgk, Yao:2020vef, Yao:2021pyf}.  Direct applicability to a wide variety of systems is a feature of the approach, which we exploit herein to deliver a coherent, parameter-free treatment of all the following semileptonic transitions: $K\to \pi$, $D\to \pi, K$ and $D_s \to K$, $B_{(s)} \to \pi(K)$; and also the related masses and leptonic decay constants.

\section{Semileptonic transitions}
\label{SecSLTransitions}
Weak semileptonic transitions are described by two scalar form factors, which express all pertinent effects of hadron structure; hence, express the impact on weak interaction observables of EHM and its modulation by Higgs boson couplings into QCD.  As an exemplar, we choose $\bar B_s^0\to K^+ \ell^- \bar\nu_\ell $.  All others can be analysed analogously.  So, consider the matrix element
{\allowdisplaybreaks
\begin{subequations}
\label{TypicalME}
\begin{align}
_u  M_\mu^{\bar B_s^0}(P,Q) & =\langle K^+(p) | \bar u i\gamma_\mu b |\bar B_s^0(k)\rangle \\
  & = f_+^{\bar B_s^u}(t) T_{\mu\nu}^Q P_\nu  + f_0^{\bar B_s^u}(t) \tfrac{P\cdot Q}{Q^2} Q_\mu\,,
\end{align}
\end{subequations}
where $Q^2T_{\mu\nu}^Q = Q^2\delta_{\mu\nu} - Q_\mu Q_\nu$,
$P =k+p$,
$Q=p-k$,
$k^2 = -m_{\bar B^0}^2$ and $p^2=-m_{K^+}^2$;
%
$t=-Q^2$;
and
$P\cdot Q  = - (m_{\bar B^0}^2 - m_{K^+}^2)$,
$P^2  = - 2 (m_{\bar B^0}^2 + m_{K^+}^2) + t$.
%
%
Symmetries entail $f_+^{\bar B_s^u}(0)= f_0^{\bar B_s^u}(0)$, ensuring the absence of kinematic singularities.
The physical domain of form factor support is limited by the masses of the hadrons involved:
writing
$t_\pm^{\bar B^0 K^+} = (m_{\bar B^0} \pm m_{K^+})^2$,
$t_-^{\bar B^0 K^+}$ is the largest $t$-value accessed in the decay process.
}

Once the form factors are known, related branching fractions can be calculated by integrating the differential decay width,
\begin{equation}
\label{dGdt}
\left.\frac{d\Gamma}{dt}\right|_{\bar B_s^0\to K^+ \ell^- \bar\nu_\ell }
= \frac{G_F^2 |V_{ub}|^2}{192\pi^3 m_{\bar B_s^0}^3}
\lambda(m_{\bar B_s^0},m_{K^+},t)^{1/2}
[1-m_\ell^2/t]^2\,
 {\mathpzc H}^2,
\end{equation}
 over $t\in [m_\ell^2,t_-^{\bar B^0 K^+}]$, where:
$G_F = 1.166 \times 10^{-5}\,$GeV$^{-2}$;
$\lambda(m_{\bar B_s^0},m_{K^+},t) = (t_+^{\bar B^0 K^+} - t)(t_-^{\bar B^0 K^+} - t)$; and
\begin{subequations}
\begin{align}
{\mathpzc H}^2 & =  [1+\tfrac{m_\ell^2}{2 t}] {\mathpzc H}_0^2+ \tfrac{3 m_\ell^2}{2t} {\mathpzc H}_t^2\,,\\
 {\mathpzc H}_0^2 & = \lambda(m_{\bar B_s^0},m_{K^+},t)\,|f_+^{\bar B_s^u}(t)|^2\,,\;
 {\mathpzc H}_t^2  =t_+^{\bar B^0 K^+} t_-^{\bar B^0 K^+} \,|  f_0^{\bar B_s^u}(t) |^2\,.
\end{align}
\end{subequations}

\section{Matrix Elements}
\label{SecRL}
The leading-order CSM approximation to $_u  M_\mu^{\bar B_s^0}$ in Eq.\,\eqref{TypicalME} is provided by the rainbow-ladder (RL) truncation \cite{Qin:2020rad}:
\begin{align}
\nonumber
_u  M_\mu^{\bar B_s^0}&(P,Q)  = 2N_c {\rm tr}\int\frac{d^4 {\mathpzc s}}{(2\pi)^4}
S_b({\mathpzc s}+k) \Gamma_{\bar B_s^0}({\mathpzc s}+k/2;k) S_s({\mathpzc s}) \\
& \times \bar\Gamma_{K^+}({\mathpzc s}+p/2;-p) S_u({\mathpzc s}+p)
i {\mathpzc W}_\mu^{ub}({\mathpzc s}+p,{\mathpzc s}+k) \,,
\label{dMDs}
\end{align}
where $N_c=3$ and the trace is over spinor indices.
Three distinct types of matrix-valued function appear in Eq.\,\eqref{dMDs}: propagators for the dressed-quarks involved in the transition process, here $S_f({\mathpzc s})$, $f=u,s,b$; Bethe-Salpeter amplitudes for the initial- and final-state mesons, $\Gamma_{M}$, $M=\bar B_s^0 , K^+$; and the dressed $b\to u$ vector weak transition vertex, ${\mathpzc W}_\mu^{ub}$.
Physically, ${\mathpzc W}_\mu^{ub}$ must exhibit poles at $Q^2 + m_{B^\ast,B_{0}^\ast}^2=0$.  They are manifest in our analysis of this transition and all analogues considered herein.

Once the kernel of the Bethe-Salpeter equation is specified, each matrix-valued function in Eq.\,\eqref{dMDs} can be calculated from the appropriate coupled integral equations, as exemplified, \emph{e.g}., in Refs.\,\cite{Ji:2001pj, Chen:2012txa, Yao:2020vef, Yao:2021pyf}.  We use the kernel specified in Refs.\,\cite{Qin:2011dd, Qin:2011xq}:
\begin{subequations}
\label{KDinteraction}
\begin{align}
\mathscr{K}_{\rho_1\rho_1',\rho_2\rho_2'} & = {\mathpzc G}_{\mu\nu}(k) [i\gamma_\mu]_{\rho_1\rho_1'} [i\gamma_\nu]_{\rho_2\rho_2'}\,,\\
 {\mathpzc G}_{\mu\nu}(k) & = \tilde{\mathpzc G}(s=k^2) T_{\mu\nu}^k\,,\\
\label{defcalG}
 \tilde{\mathpzc G}(s) & =
 \frac{8\pi^2}{\omega^4} D e^{-s/\omega^2} + \frac{8\pi^2 \gamma_m \mathcal{F}(s)}{\ln\big[ \tau+(1+s/\Lambda_{\rm QCD}^2)^2 \big]}\,,
\end{align}
\end{subequations}
where $\gamma_m=12/25$, $\Lambda_{\rm QCD} = 0.234\,$GeV, $\tau={\rm e}^2-1$, and ${\cal F}(s) = \{1 - \exp(-s/[4 m_t^2])\}/s$, $m_t=0.5\,$GeV.  
All integral equations are solved using a mass-independent chiral-limit momentum-subtraction renormalisation scheme \cite{Chang:2008ec}, with renormalisation scale $\zeta=19\,$GeV$=:\zeta_{19}$.

Explanations of Eqs.\,\eqref{KDinteraction} and their links to QCD are presented elsewhere \cite{Qin:2011dd, Qin:2011xq, Binosi:2014aea}; nevertheless, it is worth reiterating a few things here.
(\emph{a}) The kernel is consistent with that revealed by studies of QCD's gauge sector \cite{Binosi:2016xxu, Gao:2017uox, Cui:2019dwv, Kizilersu:2021jen}. 
(\emph{b}) Landau gauge is used because it supplies the conditions for which RL truncation is most accurate, being a fixed point of the renormalisation group and ensuring that sensitivity to the form of the gluon-quark vertex is minimal.
(\emph{c}) Eq.\,\eqref{defcalG} defines a two-parameter \emph{Ansatz}, $(D, \omega)$, whose details determine whether such corollaries of EHM as confinement and dynamical chiral symmetry breaking are realised in the bound-state solutions  \cite{Roberts:2020hiw, Roberts:2021nhw}.

Studies through more than two decades have shown \cite{Qin:2020rad} that interactions in the class containing Eqs.\,\eqref{KDinteraction} serve to unify the properties of many systems involving light-quarks when one uses $\omega = 0.5\,$GeV, $\varsigma^3=D\omega = (0.8\,{\rm GeV})^3$.  We employ these values.  Importantly, once $\varsigma$ is specified, then predictions for measurable quantities are largely insensitive to variations $\omega \to \omega (1\pm 0.1)$; hence, fine tuning is not an issue \cite{Qin:2020rad}.

\section{Calculation Method}
\label{SecMethod}
For systems built solely from valence degrees-of-freedom chosen from $u, d, s$ quarks (or antiquarks), every integrand involved in the calculation of quantities relevant to the transition form factors is regular throughout its integration domain.  Thus, all results can be obtained via straightforward numerical techniques.  In this way, with renormalisation group invariant light-quark current masses (in GeV)
\begin{equation}
\label{musquark}
\hat m_{u=d} = 0.0068\,, \; \hat m_s = 0.168\,,
\end{equation}
which correspond to
one-loop evolved values ($\zeta_2=2\,$GeV) $m_u^{\zeta_{2}} = 0.0044\,$GeV, $m_s^{\zeta_{2}} = 0.112\,$GeV,
one obtains the meson masses and decay constants in Table~\ref{Dstatic}A.
%
Notably, all masses and decay constants listed in Table~\ref{Dstatic} measure constructive interference between emergent and Higgs-boson mass generation \cite{Bhagwat:2006xi, Brodsky:2010xf}, \emph{e.g}., in the chiral limit, in the absence of EHM, $f_\pi \equiv 0$.

\begin{table}[t]
\caption{\label{Dstatic}
Static properties of mesons related to the transitions considered herein, calculated using the Bethe-Salpeter kernel in Eqs.\eqref{KDinteraction} and the current-quark masses in Eqs.\,\eqref{musquark}, \eqref{mcbquark}.  For comparison and where available, values recorded by the particle data group (PDG) \cite{Zyla:2020zbs} are also listed.
The mean absolute relative error between central values is 2(2)\%.
All results in Panel B were calculated using the SPM; so, each listed uncertainty expresses a $1\sigma$ confidence level on the SPM extrapolation, \emph{i.e}., 68\% of all SPM approximants give values that lie within the indicated range.
(All quantities in GeV.)
}
\begin{tabular}{l|cccc}\hline
  A  & $m_\pi$ & $f_\pi$ & $m_K$ & $f_K$  \\\hline
herein & $0.138$ & $0.093$ & $0.494$ & $0.110$  \\
PDG & $0.138$ & $0.092$ & $0.494$ & $0.110$  \\
\end{tabular}
\begin{tabular}{l|cc}\hline
  & $m_{K^\ast}$ & $m_{K^\ast_0}$ \\\hline
herein & $0.938$ & $ 0.877\phantom{(17)}$ \\
PDG & $0.892$ & $0.845(17)$ \\\hline
\end{tabular}
\medskip

\begin{tabular}{l|cccc}\hline
 B & $m_{D}$ & $f_{D}$ & $m_{D_s}$ & $f_{D_s}$ \\\hline
herein & $1.87(2)$ & $0.163(4)$ & $1.97(2)$ & $0.181(3)$ \\
PDG & $1.87\phantom{(7)}$ & $0.153(7)$ & $1.97\phantom{(7)}$ & $0.177(3)$ \\\hline
  & $m_{D^\ast}$ & $m_{D_s^\ast}$ & $m_{S_{c \bar d}}$ & $m_{S_{c\bar s}}$\\\hline
herein & $2.08(5)$ & $2.16(4)$ &$2.12(3)$  & $2.29(3)$ \\
PDG & $2.01\phantom{(5)}$ & $2.11\phantom{(5)}$ & $2.30(2)$& $2.32\phantom{(5)}$ \\\hline
  & $m_{B}$ & $f_{B}$ & $m_{B_s}$ & $f_{B_s}$\\\hline
herein & $5.26(35)$ & $0.136(17)$ &$5.39(35)$  & $0.160(14)$ \\
PDG & $5.28\phantom{(35)}$ & $0.134(01)$ & $5.37\phantom{(35)}$& $0.163(01)$ \\\hline
\end{tabular}
\begin{tabular}{l|cc}\hline
  & $m_{B^\ast}$ & $m_{B^\ast_0}$ \\\hline
herein & $5.38(36)$ & $ 5.44(36)\phantom{(17)}$ \\
PDG & $5.32\phantom{(36)}$ &  \\\hline
\end{tabular}
\end{table}

As the difference between the current-masses of the valence degrees-of-freedom in the initial-state meson increases, so does the maximum momentum-squared ($t_-$) transferred to the lepton pair.  At some value of this splitting, the growth in $t_-$ entails that singularities associated with the analytic structure of the dressed-quark propagators \cite{Maris:1997tm, Windisch:2016iud} are absorbed into the complex-${\mathpzc s}^2$ integration domain, at which point straightforward numerical integration techniques fail.
For $Q\bar{u}$, $Q\bar{s}$  initial states, this occurs at $\hat m_{Q^{\bar u}}=0.38\,$GeV, $\hat m_{Q^{\bar s}}=0.45$\,GeV, respectively.

A couple of systematic methods have been developed to overcome this limitation, \emph{viz}.\ the use of perturbation theory integral representations \cite{Chang:2013nia}, which is necessary and practicable for the calculation of form factors at large spacelike $Q^2$, and the statistical Schlessinger point method (SPM) \cite{Schlessinger:1966zz, PhysRev.167.1411, Binosi:2018rht}, based on interpolation via continued fractions, which we exploit herein.

To proceed, we consider a fictitious pseudoscalar meson $P_{Q\bar{q}}$, $q=u,s$, and compute its mass and leptonic decay constant as a function of $\hat m_Q$ up to the value $\hat m_{Q^{\bar q}}$.
Then, using the SPM, we build interpolations: $m_{P}(\hat m_Q)$, $f_{P}(\hat m_Q)$ on the domains $\hat m_Q \in [\hat m_s,m_{Q^{\bar u}}]$, to access $D$, $B$ initial states, and $\hat m_Q \in [\hat m_s,m_{Q^{\bar s}}]$, to reach $D_s$, $B_s$ mesons.
Those interpolations are then used to determine values of $\hat m_{c,b}$ such that the central (defined below) extrapolated results for $(m_{D,B} + m_{D_s,B_s})/2$ match the PDG values.  This procedure yields (in GeV):
\begin{equation}
\label{mcbquark}
\hat m_c = 1.45\,,\;
\hat m_b = 6.30\,.
\end{equation}
Using solutions of the RL gap equations defined by these renormalisation group invariant current-quark masses, one obtains the following reference values of the dressed-masses (in GeV): $M_c(\zeta_2)=1.19$, $M_b(\zeta_2)=3.90$.

It is now necessary to explain ``central'', which requires that we detail our implementation of the SPM.
In developing any interpolation, we begin with a set of results for a given quantity, ${\mathpzc Q}(\hat m_Q)$,  calculated at $N=40$ values of $\hat m_Q$ distributed evenly throughout the appropriate domain, defined above.
$M=20$ current-mass values are then chosen at random from that $40$-element set and a continued fraction interpolation developed for ${\mathpzc Q}(\hat m_Q)$ on this $20$-element subset.
One thereby obtains $C(40,20) \sim 100$-billion interpolating functions.  From that possible number, we choose the first $n_I=\,$100,000 that are singularity-free and positive definite on $\hat m_Q \in [0,2 \hat m_b]$.
Our prediction for ${\mathpzc Q}$ at any given value of $\hat m_Q>m_{Q^{\bar q}}$ on the appropriate trajectory is then obtained by extrapolating each one of the chosen $n_I$ SPM interpolants to the required current-mass and citing as the result that value located at the centre of the band within which 68\% of the interpolants' values lie.  This $1\sigma$ band is identified as the uncertainty in the result.
Such an implementation of the SPM yields the predictions in Table~\ref{Dstatic}B.

A detailed explanation of the SPM is presented elsewhere \cite{Cui:2021vgm}.  Here we simply reiterate some salient features.  The method makes no assumptions about the function used for representing input and preserves both local and global features of the source information.  The latter enhances the reliability of subsequent extrapolations.  Employing the SPM, one can accurately recover a complex-valued function within a radius of convergence fixed by that one of the function's branch points which lies nearest to the real domain that provides the sample points.  The statistical element guarantees a reliable estimate of the uncertainty associated with any extrapolation.

\begin{table}[t]
\caption{\label{SPMparameters}
Interpolation parameters for each form factor considered herein, as labelled: Eq.\,\eqref{TFfunction}, $\alpha_1$ is dimensionless and $\alpha_{2,3}$ have dimension GeV$^{-1}$.
\emph{N.B}.\ Symmetry constraints, manifest in our treatment, ensure $\alpha_1^+ = \alpha_1^0$ for all transitions; and as plain from Eq.\,\eqref{TFfunction}, $f_{+,0}(t=0) = \alpha_1^+$.
(The SPM uncertainty estimate is discussed in the paragraph after that containing Eq.\,\eqref{mcbquark}.)
}
\begin{tabular*}
{\hsize}
{
c@{\extracolsep{0ptplus1fil}}
c@{\extracolsep{0ptplus1fil}}
c@{\extracolsep{0ptplus1fil}}
c@{\extracolsep{0ptplus1fil}}
c@{\extracolsep{0ptplus1fil}}}\hline
$\alpha_1^+\ $ & $\alpha_2^+\ $ & $\alpha_3^+\ $  & $\alpha_2^0\ $ & $\alpha_3^0\ $ \\\hline
\multicolumn{5}{c}{$K^+ \to \pi^0$} \\
$0.964\phantom{(09)}\ $ & $1.156\phantom{(009)}\ $ & $1.194\phantom{(009)}\ $ & $0.723\phantom{(0009)}\ $ & $0.626\phantom{(0009)}\ $ \\ \hline
\multicolumn{5}{c}{$D^0 \to \pi^-$} \\
$0.673(09)\ $ & $0.151(05)\phantom{1}\ $ & $0.205(12)\phantom{1}\ $ & $0.0876(35)\ $ & $0.0929(92)\ $ \\ \hline
\multicolumn{5}{c}{$D_s^+ \to K^0$} \\
%
$0.681(10)\ $ & $0.180(06)\phantom{1}\ $ & $0.236(14)\phantom{1}\ $ & $0.0980(39)\ $ & $0.132(13)\phantom{1}\ $ \\ \hline
\multicolumn{5}{c}{$D^0 \to K^-$} \\
$0.796(09)\ $ & $0.180(16)\phantom{1}\ $ & $0.217(35)\phantom{1}\ $ & $0.123(18)\phantom{1}\ $ & $0.123(39)\phantom{1}\ $ \\ \hline
\multicolumn{5}{c}{$\bar B^0 \to \pi^+$} \\
$0.287(54)\ $ & $0.0130(11)\ $ & $0.0170(20)\ $ & $0.0065(08)\ $ & $0.0060(16)\ $ \\ \hline
\multicolumn{5}{c}{$\bar B_s^0 \to K^+$} \\
$0.293(55)\ $ & $0.0175(21)\ $ & $0.0182(22)\ $ & $0.0076(09)\ $ & $0.0062(17)\ $ \\ \hline
\end{tabular*}
\end{table}

\section{Transition Form Factors}
\label{SecTFFs}
Each of the form factors characterising a semileptonic transition can reliably be \emph{interpolated} using the following function:
\begin{equation}
\label{TFfunction}
f_{+,0}(\alpha_1,\alpha_2,\alpha_3,m_{\mathpzc W};t) =
 \alpha^{+,0}_1 +  \alpha^{+,0}_2 \, t+ \frac{\alpha^{+,0}_3\,  t^2}{m_{\mathpzc W^{+,0}}^2-t}\,,
\end{equation}
where $m_{\mathpzc W^{+,0}}$ is the relevant vector- or scalar-meson mass, respectively.
The calculated pole masses and interpolation coefficients for all transitions considered herein are listed in Tables~\ref{Dstatic} and \ref{SPMparameters}.  Regarding the $K^+\to\pi^0$ transition, these quantities can be calculated directly so there is no SPM uncertainty.
For the remaining transitions, the SPM is used to determine all tabulated quantities, including the interpolation coefficients, as explained following Eqs.\,(7) in Ref.\,\cite{Yao:2020vef}.
Those relating to $D$, $B$ mesons are obtained directly using this approach.  For the $D_s$, $B_s$ mesons, we instead apply the SPM to ratios of the individual coefficients computed on the $Q\bar u$ and $Q\bar s$ trajectories and multiply the associated $D$, $B$ coefficients by the extrapolation results.  This procedure capitalises on the slow variation of the ratios and returns smaller SPM uncertainties for the $D_s$, $B_s$ coefficients than direct extrapolation of coefficient values.  Within mutual uncertainties, the values obtained via either method agree.
Notably, the ratio analysis reveals that, with 99\% and 85\% confidence, respectively,
\begin{equation}
\label{EqRatios}
f_+^{D_s^d}(0)/f_+^{D_d^u}(0)>1\,, \; f_+^{\bar B_s^d}(0)/f_+^{\bar B_d^u}(0)>1\,.
\end{equation}

\subsection{$K^+ \to \pi^0$}
\begin{figure}[t]
\includegraphics[width=0.42\textwidth]{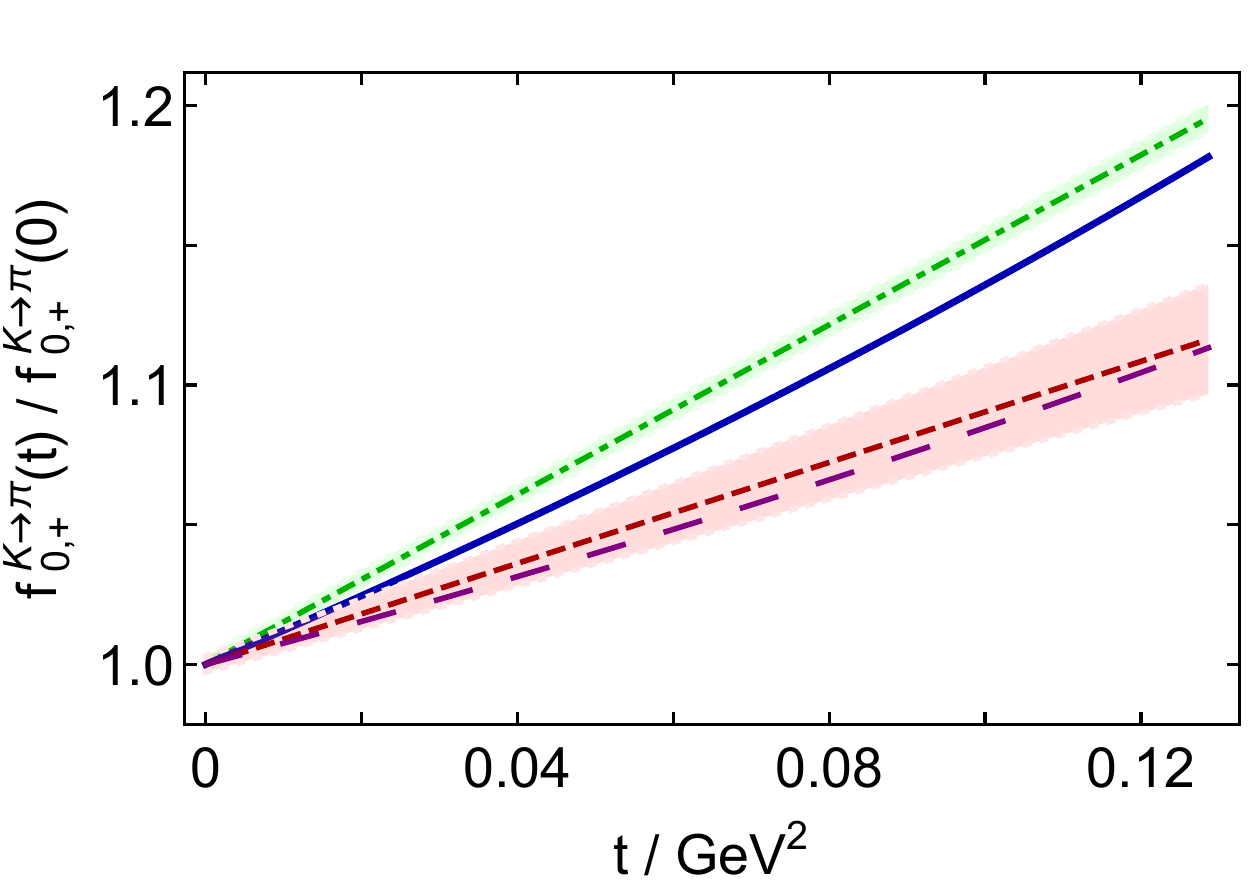}
\caption{\label{PKpi}
$K \to \pi$ transition form factors, normalised by their $t=0$ values:
$f_+^{K_u^u}$, solid blue curve; and $f_0^{K_u^u}$, long-dashed purple curve.
Comparison curves within like coloured bands are linear fits to available data described in Ref.\,\cite{Zyla:2020zbs}: $f_+^{K_u^u}$, dot-dashed green; and dashed red $f_0^{K_u^u}$.
}
\end{figure}

The $K\to\pi$ transition form factors are drawn in Fig.\,\ref{PKpi}, normalised by their $t=0$ values, for which our prediction is listed in Table~\ref{SPMparameters}: $f_+^{K_u^u}(0)=0.964=f_0^{K_u^u}(0)$.  Combined with earlier continuum results obtained using comparable methods \cite{Ji:2001pj, Chen:2012txa, Yao:2020vef}, the average is
\begin{equation}
\label{f0kpi}
f_{+,0}^{K_u^u}(0) = 0.971(9)\,,
\end{equation}
which compares well with the PDG preferred value: $0.9706(27)$.  As visible in Fig.\,\ref{PKpi}, our predicted form factors are somewhat nonlinear.  This suggests that nonlinear fitting models should be preferred in future analyses of $K\to\pi$ data.

\begin{table*}[t]
\caption{\label{TabBranch}
Pseudoscalar meson semileptonic branching fractions: each such fraction is to be multiplied by $10^{-3}$.  The column labelled ``ratio'' is the ratio of the preceding two entries in the row, so \emph{no} factor of $10^{-3}$ is applied in this column.
PDG \cite{Zyla:2020zbs} lists the following values for the CKM matrix elements:
$|V_{us}| =0.2245(8)$, $|V_{cd}| = 0.221(4)$, $|V_{cs}|= 0.987(11)$
$|V_{ub}| = 0.00382(24)$; 
and the following lifetimes (in seconds):
$\tau_{K^+}=1.2379(21)\times 10^{-8}$,
$\tau_{D^0} = 4.10 \times 10^{-13}$,
$\tau_{D_s^\pm} = 5.04 \times 10^{-13}$,
$\tau_{\bar B^0} = 1.519 \times 10^{-12}$,
$\tau_{\bar B_s^0} = 1.515 \times 10^{-12}$.
All kinematic factors are evaluated using PDG values for meson masses.
}
\begin{tabular*}
{\hsize}
{
l@{\extracolsep{0ptplus1fil}}
|c@{\extracolsep{0ptplus1fil}}
c@{\extracolsep{0ptplus1fil}}
c@{\extracolsep{0ptplus1fil}}
|c@{\extracolsep{0ptplus1fil}}
c@{\extracolsep{0ptplus1fil}}
c@{\extracolsep{0ptplus1fil}}}\hline
& \multicolumn{3}{c|}{herein} & \multicolumn{3}{c}{PDG \cite{Zyla:2020zbs} or other, if indicated} \\\hline
& $e^+ \nu_{e}\ $\phantom{[51]} & $\mu^+ \nu_\mu\ $ & ratio$\ $  & $e^+ \nu_{e}\ $ & $\mu^+ \nu_\mu\ $ & ratio$\ $ \\
$K^+\to\pi^0\ $ & $50.0(9)\phantom{00}\ $ & $33.0(6)\phantom{00}\ $ & $0.665\phantom{(70)}\ $ & $50.7(6)\phantom{00}\ $\phantom{[51]} & $33.5(3)\phantom{22}\ $ & $0.661(07)\ $ \\
$D^0\to\pi^-\ $ & $\phantom{1}2.70(12)\ $ & $\phantom{2}2.66(12)\ $ & $0.987(02)\ $ & $\phantom{0}2.91(4)\phantom{1}\ $\phantom{[51]} & $\phantom{0}2.67(12)\ $ & $0.918(40)\ $ \\
$D_s^+\to K^0\ $ & $\phantom{1}2.73(12)\ $ & $\phantom{2}2.68(12)\ $ & $0.982(01)\ $ & $\phantom{0}3.25(36)\ $\cite{Ablikim:2018upe} &  & \\
$D^0\to K^-\ $ & $39.0(1.7)\ $ & $38.1(1.7)\ $ & $0.977(01)\ $ & $\phantom{0}35.41(34)\phantom{1}\ $\phantom{[51]} & $34.1(4)\phantom{22}\ $ & $0.963(10)\ $ \\\hline
%
& $\mu^- \bar\nu_{\mu}\ $\phantom{[51]} & $\tau^- \bar\nu_\tau\ $ & ratio$\ $  & $\mu^- \bar\nu_{\mu}\ $ & $\tau^- \bar\nu_\tau\ $ & ratio$\ $ \\
$\bar B^0\to \pi^+\ $ & $\phantom{11}0.162(44)\ $ & $\phantom{11}0.120(35)\ $ & $0.733(02)\ $ & $\phantom{01}0.150(06)\ $\phantom{[51]} &  &  \\
$\bar B_s^0\to K^+\ $ & $\phantom{11}0.186(53)\ $ & $\phantom{11}0.125(37)\ $ & $0.667(09)\ $ & &  &  \\\hline
\end{tabular*}
\end{table*}

Our $K^+\to \pi^0$ semileptonic branching fractions are given in Table~\ref{TabBranch}.  They were computed using Eq.\,\eqref{dGdt}, amended to account for radiative corrections in this case \cite[Eq.\,(66.9)]{Zyla:2020zbs}, the $t=0$ result in Eq.\,\eqref{f0kpi}, and the form factors drawn in Fig.\,\ref{PKpi}.  Within quoted uncertainties, the branching fractions agree with the current PDG evaluations.

\subsection{$D^0 \to \pi^-$, $D_s^+ \to K^0$, $D^0\to K^-$}
Whilst overlapping, our predictions for transition form factors from $D_{(s)}$ initial states differ marginally from those in Ref.\,\cite{Yao:2020vef} and have smaller SPM uncertainties.  Both outcomes owe to the physical constraint of positivity that we additionally impose herein (see Sec.\,\ref{SecMethod}).

\begin{figure}[t]
\vspace*{2ex}

\leftline{\hspace*{0.5em}{\large{\textsf{A}}}}
\vspace*{-4ex}

\includegraphics[width=0.42\textwidth]{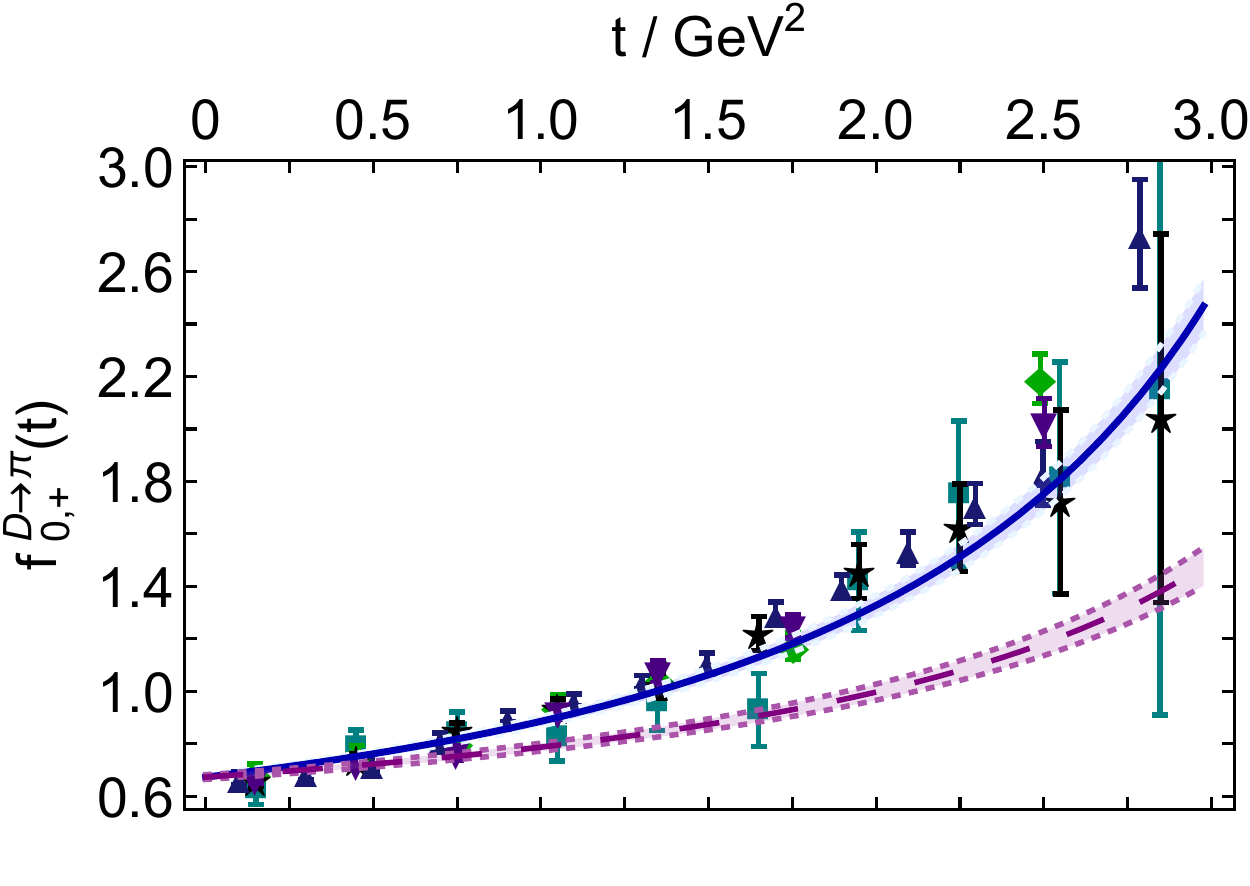}
\vspace*{-2ex}

\leftline{\hspace*{0.5em}{\large{\textsf{B}}}}
\vspace*{-4ex}
\includegraphics[width=0.42\textwidth]{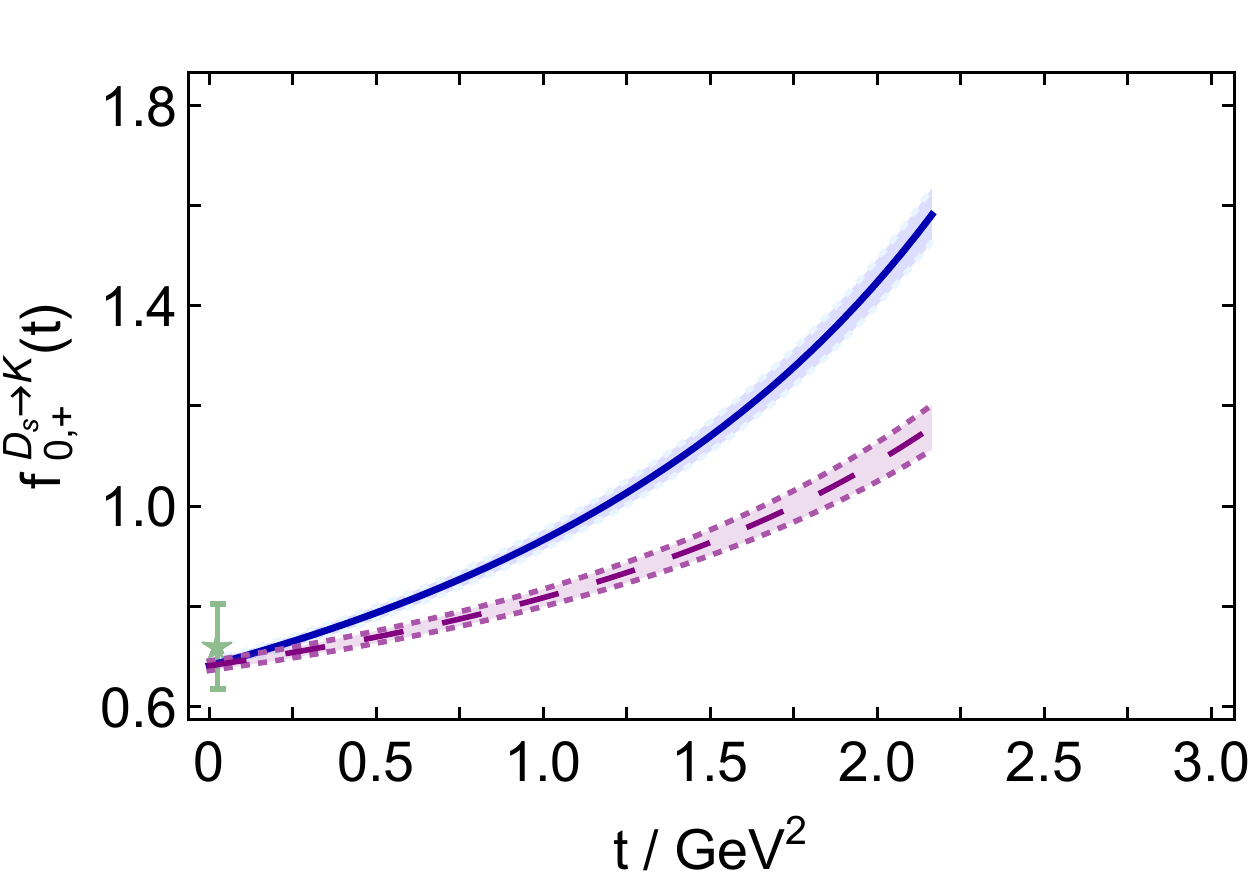}

%
\caption{\label{PDinitial}
Transition form factors from $D_{(s)}$ initial states.
\emph{Top panel}\,--\,{\sf A}.  $D^0 \to \pi^-$.
Data:
cyan squares \cite{Belle:2006idb}; green diamonds \cite{Besson:2009uv}; black stars \cite{Lees:2014ihu}; dark-blue up-triangles \cite{Ablikim:2015ixa}, and indigo down-triangles \cite{BESIII:2017ylw}.
\emph{Bottom panel}\,--\,{\sf B}.  $D_s^+ \to K^0$.  Sole available datum from Ref.\,\cite{Ablikim:2018upe}.
%
All panels: $f_+$ is depicted as the solid blue curve and $f_0$ as the long-dashed purple curve; and the like-coloured shaded bands represent the SPM uncertainty.
}
\end{figure}

The $D^0 \to \pi^-$ transition form factors are depicted in Fig.\,\ref{PDinitial}A.  Regarding $f_+^{D_u^d}$, data are available from various collaborations \cite{Belle:2006idb, Besson:2009uv, Lees:2014ihu, Ablikim:2015ixa, BESIII:2017ylw}.  The prediction is consistent with this collection, although there may be a hint of tension with the more recent sets \cite{Ablikim:2015ixa, BESIII:2017ylw} at larger values of $t$.  Owing to the presence of $\lambda(m_{D^0},m_{\pi^-},t)$ in Eq.\,\eqref{dGdt}, the contribution to the branching fraction from this domain is kinematically suppressed.

Our predictions for $D^0 \to \pi^-$ branching fractions are listed in Table~\ref{TabBranch}.
Considering first the ratio, we also computed this quantity from transition form factors obtained using a symmetry-preserving regularisation of a vector$\times$vector contact interaction \cite{Xu:2021iwv}, with the result $0.980$.  Evidently, the ratio is practically independent of the form factors used to compute the individual fractions.  Hence, given the minor impact of the $\mu$ mass on the $t$-domain accessible in the transition, the empirically determined ratio seems roughly $2\sigma$ too small.
Our result for the $\mu^+\nu_\mu$ fraction agrees with that recorded by the PDG \cite{Zyla:2020zbs}.  If this value is combined with our prediction for the ratio, then one obtains an $e^+\nu_e$ fraction of 
$0.271(10)$\%, indicating that the PDG value for this fraction is approximately $5.1\sigma$ too large when measured against its own quoted uncertainty.

Fig.\,\ref{PDinitial}B depicts our $D_s^+ \to K^0$ transition form factors.  No data are available for these form factors, except for the \mbox{$t \approx 0$} datum in Ref.\,\cite{Ablikim:2018upe}: $f_+^{D_s^d}(0) = 0.720 \pm 0.084_{\rm stat}\pm 0.013_{\rm syst}$.
Within mutual uncertainties, as evident in Fig.\,\ref{PDinitial}B, this value agrees with our prediction, listed in Table~\ref{SPMparameters}.
No $D_s^+ \to K^0$ form factor results are yet available from lattice QCD (lQCD).
Predictions for the associated branching fractions are recorded in Table~\ref{TabBranch}.  The difference between the existing measurement of the $e^+\nu_e$ final state fraction \cite{Ablikim:2018upe} and our value is $1.4\sigma$.

\begin{figure}[t]
\includegraphics[width=0.42\textwidth]{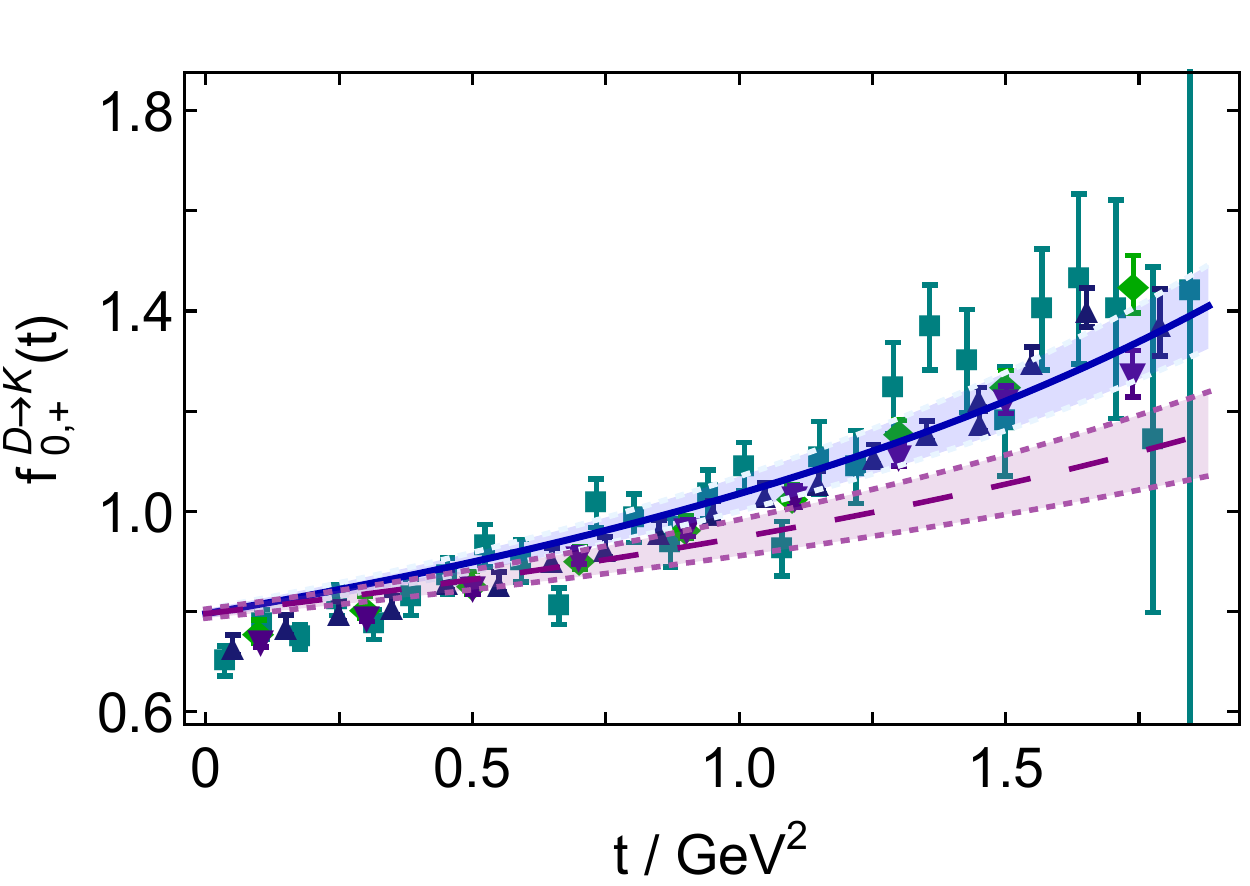}
\caption{\label{PDK}
$D\to K$ transition form factors.
$f_+$ is depicted as the solid blue curve and $f_0$ as the long-dashed purple curve; and the like-coloured shaded bands represent the SPM uncertainty.
Data:
cyan squares \cite{Belle:2006idb}; green diamonds \cite{Besson:2009uv};
dark-blue up-triangles \cite{Ablikim:2015ixa}; and indigo down-triangles \cite{BESIII:2017ylw}.
}
\end{figure}

Our $D^0\to K^-$ transition form factors are drawn in Fig.\,\ref{PDK}.  Regarding $f_+^{D_d^s}$, data are available \cite{Belle:2006idb, Besson:2009uv, Ablikim:2015ixa, BESIII:2017ylw}.  Our result is largely consistent with this collection, but there may be a hint that it sits too high at lower $t$ values.  The contributions from this domain are important to the final results for the branching fractions.  Notably, however, within mutual uncertainties, our value for $f_+^{D_d^s}(0)=0.796(9)$, listed in Table~\ref{SPMparameters}, agrees with the $N_f=2+1+1$ lQCD result reported in Ref.\,\cite{Lubicz:2017syv}: $0.765(31)$.

Our predictions for $D^0\to K^-$  branching fractions are listed in Table~\ref{TabBranch}.
Using the value of $|V_{cs}|$ listed in the caption, both the $e^+\nu_e$ and $\mu^+\nu_\mu$ fractions exceed their respective PDG values; but the ratio agrees within $1.4\sigma$, which indicates that a common overall factor should remedy the mismatch.
Adopting this perspective, one finds that the value $|V_{cs}| = 0.937(17)$
combined with our form factors delivers branching fractions in agreement with the PDG values, \emph{viz}.\ $3.52(18)$\% and $3.44(18)$\%, respectively.
In fact, referring to Ref.\,\cite[Sec.\,12.2.4]{Zyla:2020zbs}, one sees that our inferred value
is both a match for and more precise than one of the two used by the PDG to arrive at the average listed in the caption of Table~\ref{TabBranch}.  Using $|V_{cs}| = 0.937(17)$ instead to  compute this average, one finds
\begin{equation}
\label{VcspredictionUpdate}
|V_{cs}| = 0.974(10)\,.
\end{equation}

\subsection{$\bar B^0 \to \pi^+$, $\bar B_s^0 \to K^+$}
The Introduction highlighted that a sound theoretical treatment of these transitions is challenging because of the Nambu-Goldstone boson character of pions and kaons and the large disparity between the masses of the initial and final states.  CSMs have proven particularly effective in dealing with the former \cite{Maris:1997hd, Brodsky:2010xf, Chang:2013pq, Horn:2016rip, Qin:2020jig}.  Regarding the latter, it was highlighted elsewhere \cite{Xu:2021iwv} that in order to achieve a good description of such transitions on the entire kinematic domain, a realistic representation of the quark+antiquark interaction at large relative momentum is necessary.  Our approach expresses this; and with its capacity for delivering precise results for transition form factors on $\hat m_Q \in [0,\hat m_{Q^{\bar s}}]$, then, augmented by the SPM, our CSM formulation also circumvents the mass disparity issue.

\begin{figure}[t]
\vspace*{2ex}

\leftline{\hspace*{0.5em}{\large{\textsf{A}}}}
\vspace*{-4ex}

\includegraphics[width=0.42\textwidth]{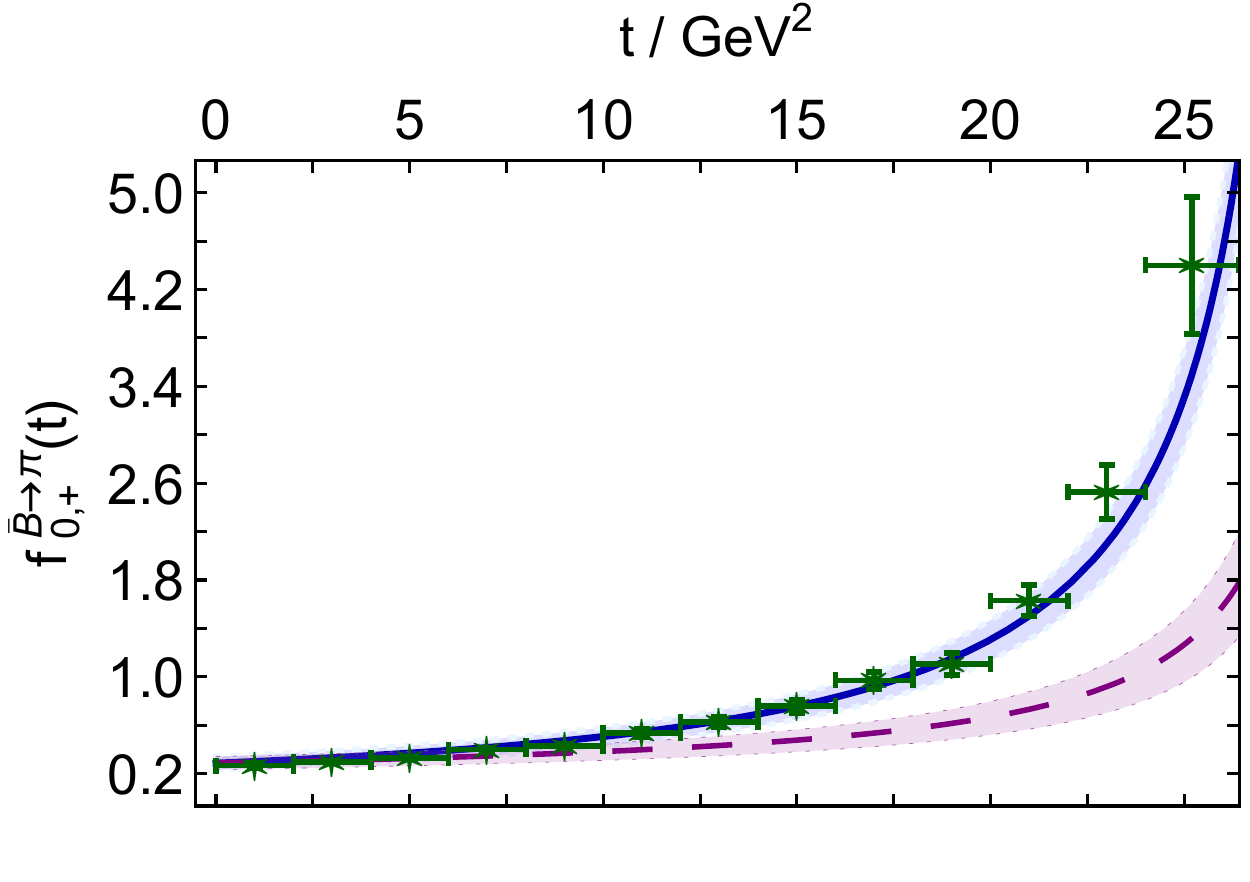}
\vspace*{-2ex}

\leftline{\hspace*{0.5em}{\large{\textsf{B}}}}
\vspace*{-4ex}
\includegraphics[width=0.42\textwidth]{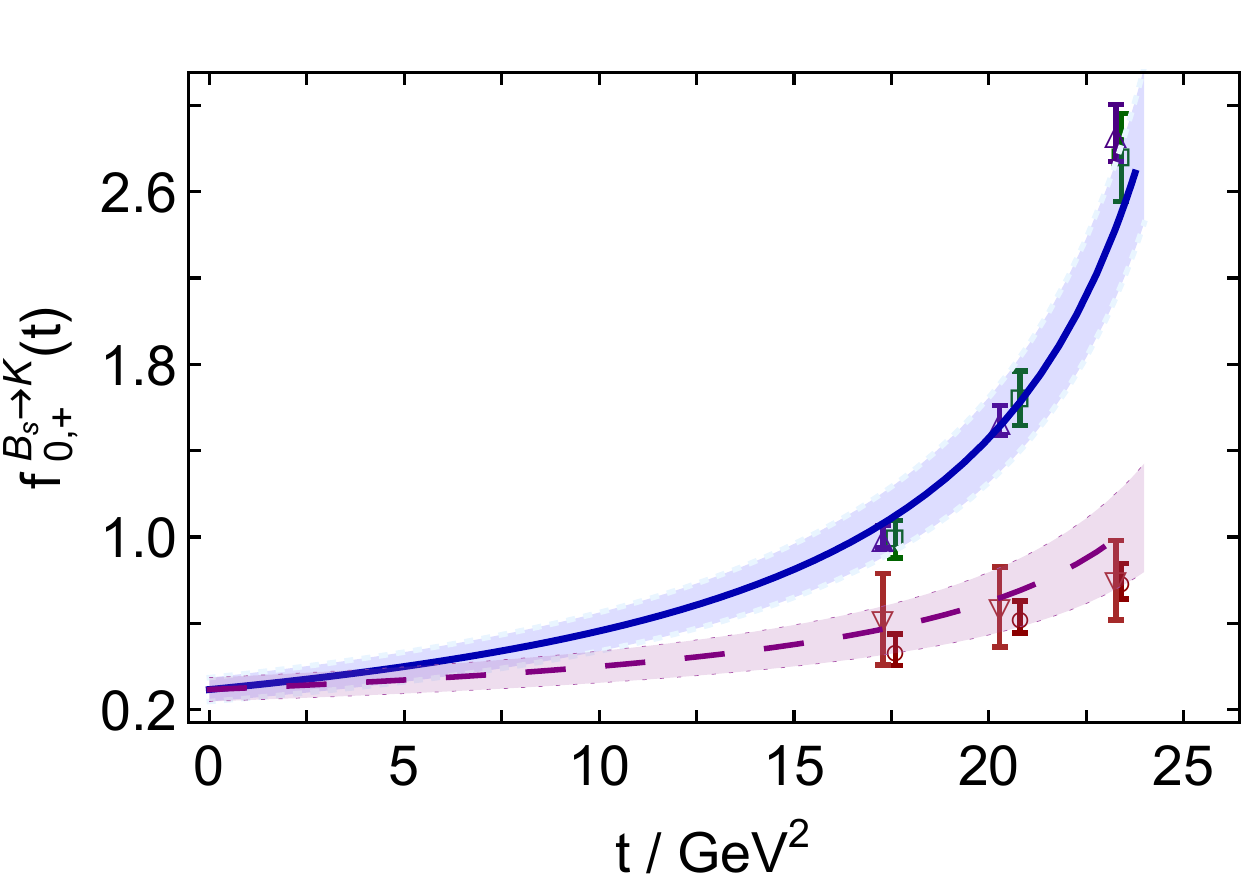}

%
\caption{\label{PBinitial}
$B_{(s)}$ transition form factors: $f_+$ -- solid blue curve; $f_0$ -- long-dashed purple curve; and the like-coloured shaded bands represent the SPM uncertainty.
\emph{Top panel}\,--\,{\sf A}.  $\bar B^0 \to \pi^+$.
Data:
Data (green stars): reconstructed
from the average \cite[Tab.\,81]{HFLAV:2019otj} of data reported in Refs.\,\cite{BaBar:2010efp, Belle:2010hep, BaBar:2012thb, Belle:2013hlo}.
\emph{Bottom panel}\,--\,{\sf B}.  $\bar B_s^0 \to K^+$.  lQCD results:
$f_+$ -- indigo open up-triangles \cite{Bouchard:2014ypa} and green open boxes \cite{Flynn:2015mha};
$f_0$ --- brown open down-triangles \cite{Bouchard:2014ypa} and red open circles \cite{Flynn:2015mha}.
}
\end{figure}

Our results for the $\bar B^0 \to \pi^+$ transition form factors are drawn in Fig.\,\ref{PBinitial}A.  Concerning $f_+^{\bar B_u^d}$, data have been collected by various collaborations \cite{BaBar:2010efp, Belle:2010hep, BaBar:2012thb, Belle:2013hlo}; and within mutual uncertainties, our predictions agree with this data.  The data supports a value
\begin{equation}
f_+^{\bar B_u^d}(t=0) = 0.27(2)\,,
\end{equation}
which is consistent with our prediction, Table~\ref{SPMparameters}: $0.29(5)$.

Using the form factors in Fig.\,\ref{PBinitial}A, we obtain the $\bar B^0 \to \pi^+$ branching fractions in Table~\ref{TabBranch}.  The PDG lists a results for the $\mu^-\nu_\mu$ final state, which matches our prediction within mutual uncertainties.  Exact agreement is obtained using
\begin{equation}
|V_{ub}| = 0.00374(44)\,.
\end{equation}
This value is consistent with other analyses of ${\mathpzc B}_{\bar B^0\to \pi^+ \mu^-\bar\nu_\mu}$ \cite[Sec.\,75.3]{Zyla:2020zbs}; hence, increases tension with the higher value inferred from inclusive decays.
No empirical information is available on the $\tau^-\nu_\tau$ final state and, consequently, the $\tau$:$\mu$ ratio.  Here, a $N_f=2+1$-flavour lQCD study yields $0.69(19)$ \cite{Flynn:2015mha}, which, within its uncertainty, matches our result:  $0.733(2)$.

Fig.\,\ref{PBinitial}B displays our calculated $\bar B_s\to K^+$ form factors.  Although the $B_s \to K^-$ transition has recently been observed \cite{LHCb:2020ist}, no form factor data are yet available.  We therefore compare our predictions with results obtained using  $N_f=2+1$-flavour lQCD \cite{Bouchard:2014ypa, Flynn:2015mha}.
Owing to the difficulties attendant upon use of lattice methods to calculate form factors involving heavy+light mesons, lQCD results are restricted to a few points on the domain $t \gtrsim 17\,$GeV$^2$, as seen in Fig.\,\ref{PBinitial}B.  lQCD analyses now typically employ such results to develop a least-squares fit to the form factor points, using a practitioner-preferred functional form.  That fit is subsequently employed to define the form factor on the entire kinematically accessible domain, $0\lesssim t \lesssim 25\,$GeV$^2$.
At this time, given the paucity of points and their limited precision, the SPM cannot profitably be used to develop function-form unbiased representations of the lQCD output.

\begin{figure}[t]
\vspace*{2ex}

\leftline{\hspace*{0.5em}{\large{\textsf{A}}}}
\vspace*{-4ex}

\includegraphics[width=0.42\textwidth]{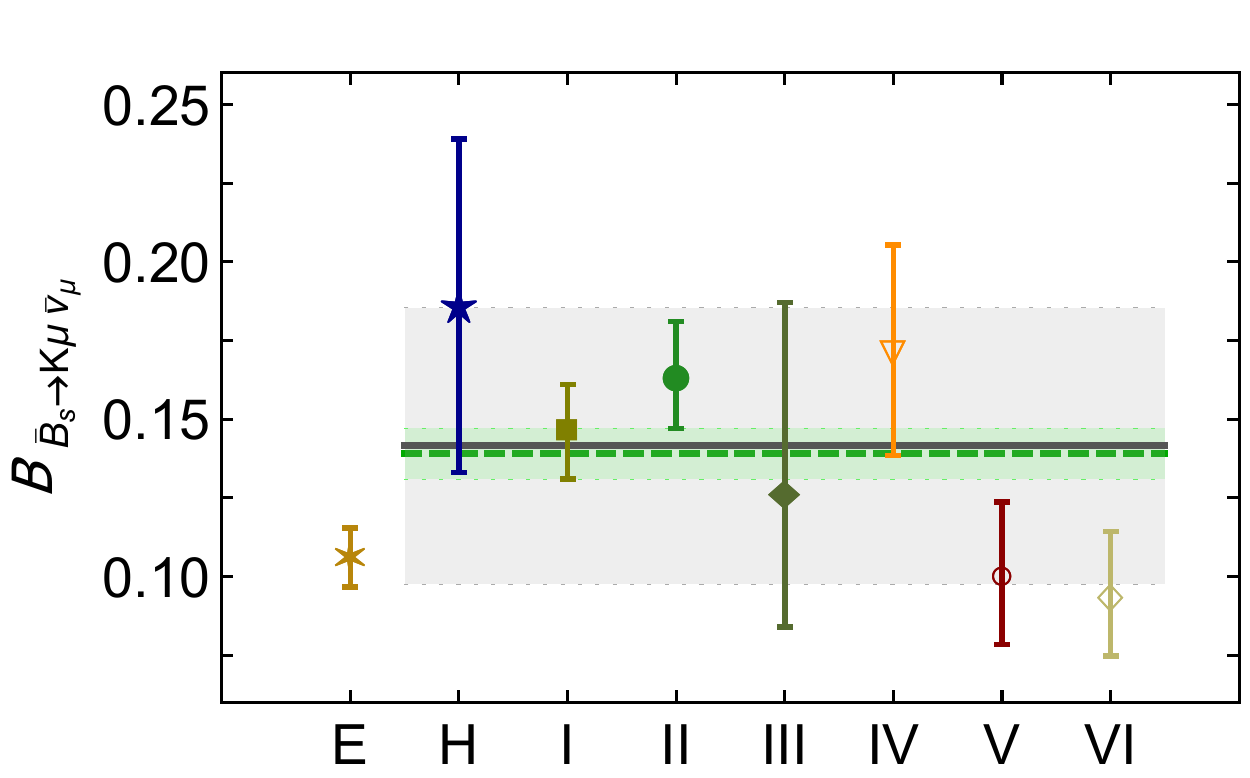}

\leftline{\hspace*{0.5em}{\large{\textsf{B}}}}
\vspace*{-4ex}
\includegraphics[width=0.42\textwidth]{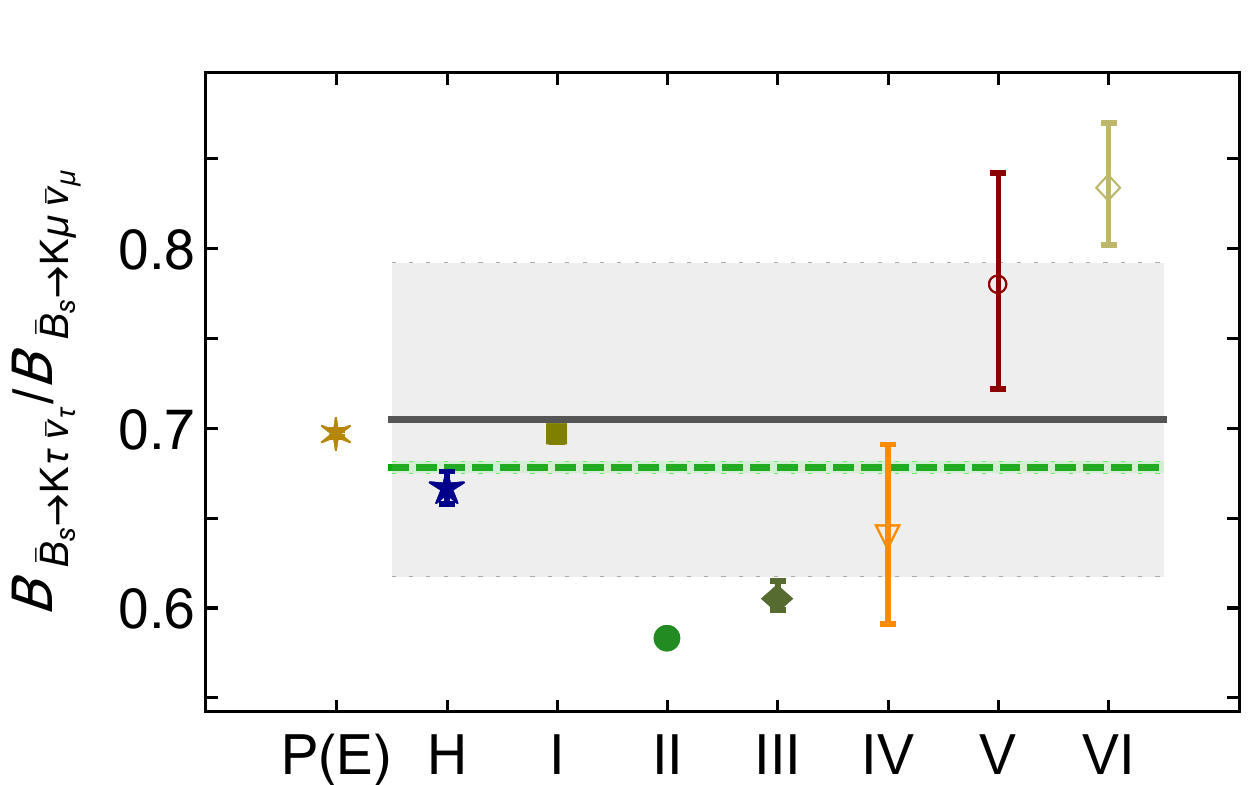}
%
\caption{\label{PBFBK}
\emph{Top panel}\,--\,{\sf A}.
Branching fraction ${\mathpzc B}_{\bar B_s^0\to K^+ \mu^- \bar\nu_\mu}$ computed herein, ``H'', compared with
the value in Eq.\,\eqref{LHCbmeasure}, ``E'', \emph{viz}.\ a measurement of ${\mathpzc B}_{B_s^0\to K^- \mu^+ \nu_\mu}$ \cite{LHCb:2020ist}
and some results obtained using other means: continuum I -- III \cite{Wu:2006rd, Faustov:2013ima, Xiao:2014ana}; and lattice IV -- VI \cite{Bouchard:2014ypa, Flynn:2015mha, FermilabLattice:2019ikx}.
%
%
\emph{Bottom panel}\,--\,{\sf B}.
Branching fraction ratio ${\mathpzc B}_{\bar B_s^0\to K^+ \tau^- \bar\nu_\tau}/{\mathpzc B}_{\bar B_s^0\to K^+ \mu^- \bar\nu_\mu}$ computed herein compared with some results obtained using other means.  Legend as in Panel~{\sf A}, except ``P(E)'' is the result from \cite{Gonzalez-Solis:2021awb}, which is an estimate constrained by the datum in Ref.\,\cite{LHCb:2020ist}.
Both panels -- Grey line and like-coloured band: unweighted mean of theory results and related uncertainties.
Green dashed line and green band: uncertainty-weighted average of theory results and associated uncertainty.
}
\end{figure}

The form factors in Fig.\,\ref{PBinitial}B produce the $\bar B_s^0 \to K^+$ branching fractions in Table~\ref{TabBranch}.  Ref.\,\cite{LHCb:2020ist} reports the following value:
\begin{equation}
\label{LHCbmeasure}
{\mathpzc B}_{B_s^0\to K^- \mu^+ \nu_\mu} = [ 0.106 \pm 0.005_{\rm stat} \pm 0.008_{\rm syst}] \times 10^{-3}\,.
\end{equation}
Fig.\,\ref{PBFBK} compares this measurement with our prediction and also results obtained via various other means: experiment and theory are in agreement, but only because the theory uncertainty is large.
The unweighted average of the theoretical results is
$0.141(44)\times 10^{-3}$ and the uncertainty weighted mean is $0.139(08)\times 10^{-3}$.
Omitting entries V--VI \cite{Flynn:2015mha, FermilabLattice:2019ikx}, these results increase:
unweighted $0.159(38)\times 10^{-3}$ and uncertainty weighted $0.156(10)\times 10^{-3}$.
The extrapolations employed in Refs.\,\cite{Flynn:2015mha, FermilabLattice:2019ikx} lead to values of $f_+^{B_s^u}(0)$ that are roughly one-half of those obtained in I--IV \cite{Wu:2006rd, Faustov:2013ima, Xiao:2014ana, Bouchard:2014ypa}: $0.148(53)$  vs.\ $0.299(86)$.  This is sufficient to explain the difference in branching fractions: V--VI vs.\ I--IV in Fig.\,\ref{PBFBK}.
An alternative approach to fitting and extrapolating lQCD results, using the datum in Ref.\,\cite{LHCb:2020ist} as an additional constraint, produces \cite{Gonzalez-Solis:2021awb}: $f_+^{\bar B_s^u}(0)=0.211(3)$.
%
We note that Ref.\,\cite{Gonzalez-Solis:2021awb} also infers $f_+^{\bar B_d^u}(0)=0.255(5)$, leading to $f_+^{\bar B_s^d}(0)/f_+^{\bar B_d^u}(0)<1$.  This conflicts with our analysis -- Eq.\,\eqref{EqRatios}, and the results in a raft of other studies, \emph{e.g}., Refs.\,\cite{Melikhov:2001zv, Faessler:2002ut, Ebert:2003wc, Ball:2004ye, Wu:2006rd, Khodjamirian:2006st, Lu:2007sg, Ivanov:2007cw, Faustov:2013ima}.  Hence, the Ref.\,\cite{Gonzalez-Solis:2021awb} value for $f_+^{\bar B_s^u}(0)$ may be too small.

No empirical information is available on the $\tau \nu_\tau$ final state; hence, none on the $|V_{ub}|$-independent ratio that would test lepton flavour universality.  So, in Fig~\ref{PBFBK}B, we compare our prediction for this ratio, listed in Table~\ref{TabBranch}, with results obtained via other means.
The unweighted average of the theory results is
$0.705(87)$ and the uncertainty weighted mean is $0.678(03)$.
Omitting entries V--VI \cite{Flynn:2015mha, FermilabLattice:2019ikx}, these results are:
unweighted $0.653(41)$ and uncertainty weighted $0.677(03)$.
Within internally consistent analyses, many uncertainties cancel in this ratio; so the values should be more reliable than a calculation of either fraction alone.  Hence, the scatter in results indicates that there is ample room for the precision of $\bar B_s^0 \to K$ theory to improve.

\section{Conclusions and Perspectives}
We employed a systematic, symmetry-preserving approach to the continuum strong-interaction bound-state problem to deliver a unified set of parameter-free predictions for
[Table~\ref{Dstatic}] the leptonic decay constants of $\pi$, $K$, $D_{(s)}$, $B_{(s)}$ mesons,
and [Sec.\,\ref{SecTFFs}] the semileptonic $K\to \pi$, $D\to \pi, K$ and $D_s \to K$, $B_{(s)} \to \pi(K)$ transition form factors, each on their entire physical kinematic domain,
and the derived branching fractions.
%
%
Key results  include
an indication that the PDG value for ${\mathpzc B}_{D^0\to \pi^- e^+ \nu_e}$ is $5.1\sigma$ too large, when measured against its own quoted uncertainty;
quantitative agreement with all measured transition form factors;
improved precision on the value of $|V_{cs}|$ from semileptonic transitions;
predictions for the hitherto unmeasured $D_s\to K^0$, $\bar B_s \to K^+$ form factors;
and predictions for five as yet unmeasured branching fractions and all branching fraction ratios in this sector that can be used to test lepton flavour universality.

Notably, too, we present a quantitative analysis that compares extant theory and the recent measurement of ${\mathpzc B}_{B_s^0\to K^- \mu^+ \nu_\mu}$.   Broadly, there is agreement; but the lack of precision in contemporary theory prevents any firm conclusions.   Measurements enabling extraction of $B_s^0\to K^-$ transition form factors would be especially useful for refining both comparisons with theory and between theory analyses; hence, making progress toward a more accurate value of $|V_{ub}|$.

Efforts are underway to improve the precision of our results.  Moreover, the extension to semileptonic transitions with vector-meson final states is almost complete.  Such a comprehensive, unified set of parameter-free predictions for leptonic decay constants and heavy-to-light semileptonic transitions will enable deeper and wider understanding of the roles played by emergent hadron mass, Higgs-boson couplings, and the interference between them in forming observable phenomena; especially, their effects on the values of the CKM matrix elements.

\smallskip

\noindent\textbf{Acknowledgments}\,---\,
%
We are grateful for constructive comments from X.-W.~Kang, Y.-M.~Wang and Z.-N.~Xu.
Work supported by:
National Natural Science Foundation of China (grant nos.\,12135007 and 11805097);
Jiangsu Provincial Natural Science Foundation of China (grant no.\,BK20180323);
Nanjing University Innovation Programme for PhD candidates (grant no.\,CXYJ21-29);
and
STRONG-2020 ``The strong interaction at the frontier of knowledge: fundamental research and applications'', which received funding from the European Union’s Horizon 2020 research and innovation programme under grant agreement No 824093.
%





\end{document}